\documentclass[english,submit]{ipsj}

\usepackage[cmex10]{amsmath}
\usepackage{latexsym}
\usepackage{graphicx}
\usepackage{url}
\usepackage{color}
\usepackage{bm}

\def\Underline{\setbox0\hbox\bgroup\let\\\endUnderline}
\def\endUnderline{\vphantom{y}\egroup\smash{\underline{\box0}}\\}
\def\|{\verb|}

\usepackage[varg]{txfonts}
\makeatletter%
\input{ot1txtt.fd}
\makeatother%

\begin{document}

\title{Hybrid Numerical Solvers for Massively Parallel Eigenvalue Computation
and Their Benchmark with Electronic Structure Calculations}

\affiliate{TUandCREST}{Tottori University, JST-CREST}

\author{Hiroto Imachi}{TUandCREST}[D14T1001B@edu.tottori-u.ac.jp]
\author{Takeo Hoshi}{TUandCREST}[]

\begin{abstract}

  Optimally hybrid numerical solvers were constructed
  for massively parallel generalized eigenvalue problem (GEP).
  The strong scaling benchmark was carried out on the K computer and other supercomputers
  for electronic structure calculation problems in the matrix sizes of  $M=10^4-10^6$
  with upto $10^5$ cores.
  The procedure of GEP  is decomposed into
  the two subprocedures  of
  the reducer to the standard eigenvalue problem (SEP) and
  the solver of SEP.
  A hybrid solver is constructed,
  when a routine is chosen for each subprocedure
  from the three parallel solver libraries of ScaLAPACK, ELPA and EigenExa.
   The hybrid solvers with the two newer libraries, ELPA and EigenExa, give
  better benchmark results than the conventional ScaLAPACK library.
  The detailed analysis on the results implies that
  the reducer can be a bottleneck in next-generation (exa-scale) supercomputers,
  which indicates the guidance for future research.
  The code was developed as a middleware and a mini-application and
  will appear online.

\end{abstract}


\maketitle

\section{Introduction}
\label{sec:Introduction}
Numerical linear algebraic solvers for large matrices
have strong needs among various applications
with the current and next-generation supercomputers.
Nowadays
ScaLAPACK\cite{ScaLAPACK, SCALAPACK-URL}
\footnote{
  ScaLAPACK = Scalable Linear Algebra PACKage
} is the {\it de facto} standard solver library for parallel computations
but several routines give severe bottlenecks
in the computational speed with current massively parallel architectures.
Novel solver libraries were proposed so as to overcome the bottlenecks.
Since the performance of numerical routines
varies significantly with problems and architectures,
the best performance is achieved,
when one constructs an optimal \lq hybrid' among the libraries.

\begin{figure}[htb]
  \centering
  \includegraphics[width=7cm]{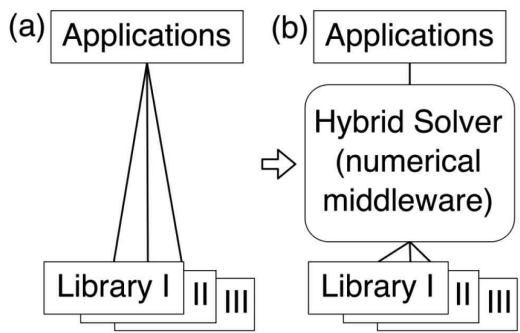}
  \caption{Concept of hybrid solver; Structure of the program code
    (a) without and (b) with hybrid solver or numerical middleware.}
  \label{fig:Concept}
\end{figure}

The concept of hybrid solver
is illustrated in Fig.~\ref{fig:Concept}.
It is a numerical middleware
and has a unique data interface to real applications.
One can choose the optimal workflow
for each problem without any programming effort.

The present paper focuses on
dense-matrix solvers for
generalized eigenvalue problems (GEPs)
in the form of
\begin{eqnarray}
  A \bm{y}_k = \lambda_k B \bm{y}_k
  \label{EQ-GEV-EQ}
\end{eqnarray}
with the given $M \times M$ real-symmetric matrices of $A$ and $B$.
The matrix $B$ is positive definite.
The eigenvalues $\{ \lambda_k \}$
and the eigenvectors $\{ \bm{y}_k \}$ will be calculated.
The computational cost is
$\mathcal{O}(M^3)$ or is proportional to $M^3$.
The present hybrid solvers are constructed
among ScaLAPACK and the two newer libraries of
ELPA
~\cite{ELPA-URL, ELPAReview, ELPAAlgorithm}
\footnote{
  ELPA = Eigenvalue soLvers for Petascale Applications
},
and EigenExa~\cite{EIGENEXA-URL, EigenExa-PNST, EigenExa-PMAA, EigenExa-ISC}.
The ELPA and EigenExa libraries are written in Fortran and
appeared in 2000's
for efficient massively parallel computations.

The present paper is organized as follows;
Section ~\ref{SEC-BACKGROUND} explains the background
from the electronic structure calculation.
Section ~\ref{SEC-WORKFLOW} describes
the mathematical foundation.
Sections ~\ref{SEC-BENCHMARK} and
~\ref{SEC-DISCUSSIONS} are
devoted to the benchmark results and discussions, respectively.
The summary and future outlook
will appear in Sec.~\ref{SEC-SUMMARY}.

\begin{figure}[htb]
  \centering
  \includegraphics[width=6cm]{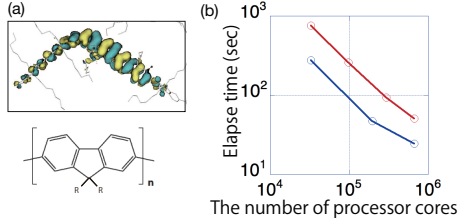}
  \caption{(a) The upper panel is a $\pi$-type electronic wavefunction in an amorphous-like
    conjugated polymer (poly-((9,9) dioctyl fluorine)).
    The lower panel shows the atomic structure (R$\equiv$C$_8$H$_{17}$) \cite{HOSHI2014-JPS-CP}.
    (b) Strong scaling plot by ELSES for one-hundred-million-atoms calculations on the K computer. \cite{HOSHI2014-JPS-CP,HOSHI-TAIWAN-PROC} The calculated materials are
    a nano-composite carbon solid (the upper line) and the amorphous-like conjugated polymer (the lower line).
    The number of used processor nodes are from $P=$ 4,096 to 82,944 (full nodes of the K computer).
    \label{fig:graph-bench-100M-aPF-NCCS}}
\end{figure}

\section{Background \label{SEC-BACKGROUND}}

\subsection{Large-scale electronic structure calculations}

The GEP of Eq.~(\ref{EQ-GEV-EQ})
gives the mathematical foundation
of electronic structure calculations
or quantum mechanical calculations of materials,
in which
an electron is treated as a quantum mechanical \lq wave'.
The input matrix $A$ or $B$ of Eq.~(\ref{EQ-GEV-EQ})  is
called Hamiltonian or the overlap matrices, respectively.
An eigenvalue of $\{ \lambda_k \}$ is
the energy of one electron and an eigenvector of
$\{ \bm{y}_k \}$ specifies the wavefunction
or the shape of an electronic \lq wave'.
Fig. \ref{fig:graph-bench-100M-aPF-NCCS}(a)
shows an example of the wavefunction.
The number of the required eigenvalues
is, at least, on the order of the number of
the electrons or the atoms in calculated materials.
See the ELPA paper ~\cite{ELPAReview} for a review,
because ELPA was developed under
tight collaboration with electronic structure calculation society.

Here, our motivation is explained.
The present authors developed
a large-scale quantum material simulator
called ELSES
\footnote{
  ELSES = Extra-Large-Scale Electronic Structure calculation
}
\cite{ELSES-URL,HOSHI-mArnoldi}.
The theories are explained
in Refs.~\cite{HOSHI-mArnoldi, SOGABE-2012-GSQMR} and the reference therein.
The matrices are based on
the real-space atomic-orbital representation and
the matrix size $M$ is
nearly proportional to the number of atoms $N$
($M \propto N$).
The simulations mainly use
novel \lq order-$N$' linear-algebraic methods
in which the computational cost
is \lq order-$N$' ($\mathcal{O}(N)$)
or is proportional to the number of atoms $N$.
Their mathematical foundation is
sparse-matrix (Krylov-subspace) solvers.
Efficient massively parallel computation is found in
Fig.~\ref{fig:graph-bench-100M-aPF-NCCS},
a strong scaling benchmark on the K computer
\cite{HOSHI2014-JPS-CP, HOSHI-TAIWAN-PROC}
with one hundred million atoms or
one-hundred-nanometer scale materials.
The simulated materials are
a nano-composite carbon solid with $N=103,219,200$
or $M=412,876,800$
\cite{HOSHI2014-JPS-CP}
and an amorphous-like conjugated polymer
with $N=102,238,848$ or $M=230,776,128$
\cite{HOSHI-TAIWAN-PROC}.

The present dense-matrix solvers
are complementary methods to the order-$N$ calculations,
because the order-$N$ calculation gives approximate solutions,
while the dense-matrix solvers give
numerically exact ones
with a heavier ($\mathcal{O}(M^3)$) computational cost.
The use of the two methods will lead us
to fruitful researches.
The exact solutions are important, for example,
when the system has many nearly degenerated eigen pairs
and one would like to distinguish them.
The exact solutions are important also as reference data
for the development of fine approximate solvers.

The  matrices of $A$ and $B$ in the present benchmark appear on
\lq ELSES Matrix Library'.~\cite{ELSESMatrixLibrary}
The Library is the collection of
the matrix data generated by ELSES for material simulations.
The benchmark was carried out
with the data files of
\lq NCCS430080', \lq VCNT22500'
\lq VCNT90000' and \lq VCNT1008000'
for the matrix sizes of
$M$=22,500, $M$=90,000, $M$=430,080,
$M$=1,008,000, respectively.
A large matrix data ($>0.5$GB) is uploaded
as a set of split files for user's convenience.

The physical origin of the matrices is explained briefly.
The files in the present benchmark are carbon materials
within modeled tight-binding-form theories based on {\it ab initio} calculations.
The matrix of \lq NCCS430080'
appears in our material research
on a nano-composite carbon solid (NCCS)
\cite{HOSHI-NCCS}.
An sp-orbital form \cite{CALZAFERRI} is used and
the system contains $N=M/4=107,520$ atoms.
The other files are generated
for thermally vibrated single-wall carbon nanotubes (VCNTs) within a supercell.
An spd-orbital form \cite{CERDA2000} is used and
each system contains $N=M/9$ atoms.
The VCNT systems were prepared, so as to generate matrices
systematically in different size with similar eigenvalue distributions.
We used these matrices
for the investigation on $\pi$-electron materials
with the present dense-matrix solver and the order-$N$ solver.
\footnote{
The present matrices are sparse,
which does not lose the generality of the benchmark,
since the cost of the dense matrix solver is not dependent
on the number of non-zero elements of the matrix.
}

\begin{figure}[htb]
  \centering
  \includegraphics[width=6cm]{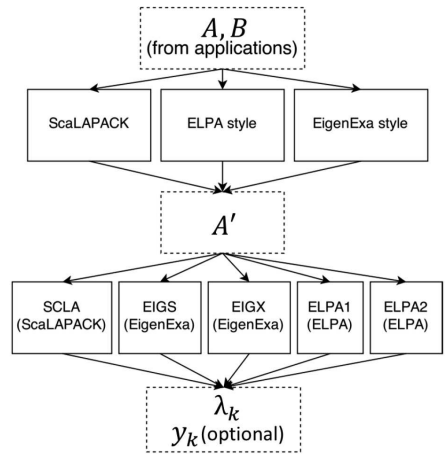}
  \caption{Workflow of the hybrid GEP solver.}
  \label{fig:EigenTestDetailedFlow}
\end{figure}

\section{The hybrid solvers \label{SEC-WORKFLOW}}

A hybrid solver is constructed,
when a routine is chosen
for each subprocedure from ScaLAPACK, EigenExa and ELPA.
The code was developed as a general middleware
that can be connected
not only to ELSES but also to any real application software,
as in Fig.~\ref{fig:Concept}.
A mini-application was also developed and
used in the present benchmark.
In the benchmark,
ScaLAPACK was used as a built-in library on each machine.
EigenExa in
the version 2.2a
\footnote{
  The present EigenExa package does not include the GEP solver.
  The GEP solver routine for EigenExa in the present paper
  is that of the version 2.2b of KMATH\_EIGEN\_GEV
  \cite{EigenExaGEV} that
  shares the SEP solver routine with the EigenExa package.
}
and ELPA in the version 2014.06.001 were used.
ELPA and EigenExa call some ScaLAPACK routines.

\subsection{Mathematical formulation \label{SEC-MATH-FOUND}}

The GEP of Eq.~(\ref{EQ-GEV-EQ})
can be written in a matrix form of
\begin{eqnarray}
  A Y = BY \Lambda,
  \label{EQ-GEV-EQ-MAT}
\end{eqnarray}
where the matrix $\Lambda \equiv {\rm diag}(\lambda_1, \lambda_2, \dots)$ is diagonal and
the matrix $Y \equiv (\bm{y}_1 \, \bm{y}_2 \, \cdots)$ satisfies $Y^T B Y = I$.
In the solvers,
the GEP of Eq.~(\ref{EQ-GEV-EQ})
is reduced to a standard eigenvalue problem (SEP) of
\begin{eqnarray}
  A' Z = Z \Lambda,
  \label{EQ-SEV-EQ}
\end{eqnarray}
where the reduced matrix $A'$ is real symmetric
\cite{MatrixComputations}
and the matrix of $Z \equiv (\bm{z}_1 \, \bm{z}_2 \, \cdots)$ contain eigenvectors of $A'$.
The reduction procedure can be achieved,
when the Cholesky factorization of $B$ gives
the Cholesky factor $U$ as an upper triangle matrix:
\begin{eqnarray}
  B = U^T U.
  \label{EQ-CHOLE-DECMP}
\end{eqnarray}
The reduced matrix $A'$ is defined by
\begin{eqnarray}
  A' = U^{-T} A U^{-1}.
  \label{EQ-A-ATAU}
\end{eqnarray}
The eigenvectors of the GEP,
written as $Y \equiv (\bm{y}_1 \, \bm{y}_2 \, \cdots)$,
are calculated
from those of the SEP by
\begin{eqnarray}
  Y = U^{-1} Z.
  \label{EQ-BACKWARD-TRANS}
\end{eqnarray}
This procedure is usually called backward transformation.

The GEP solver is decomposed into the two subprocedures of
(a) the solver of the SEP in Eq.~(\ref{EQ-SEV-EQ}) and
(b) the reduction from the GEP to the SEP
($( A, B ) \Rightarrow A'$)
and the backward transformation
($Z \Rightarrow Y$).
The subprocedures (a) and (b) are called
\lq SEP solver' and \lq reducer', respectively, and
require $\mathcal{O}(M^3)$ operations.

Figure ~\ref{fig:EigenTestDetailedFlow} summarizes the workflows
of the possible hybrid solvers.
A hybrid solver is constructed,
when one choose the routines for (a) the SEP solver
and (b) the reducer, respectively.

For (a) the SEP solver,
five routines are found in the base libraries;
One routine is a ScaLAPACK routine
(routine name in the code : \lq pdsyevd')
that uses the conventional tridiagonalization algorithm.
\cite{REF-pdsyevd}
The ELPA or EigenExa library contains
a SEP solver routine based on  the tridiagonalization algorithm.
The routine in ELPA is
called \lq ELPA1' (routine name in the code : \lq solve\_evp\_real') in this paper,
as in the original paper \cite{ELPAReview},
and the one in EigenExa
called \lq Eigen\_s' or \lq EIGS' (routine name in the code : \lq eigen\_s').
ELPA and EigenExa also contain
the novel SEP solvers based on
the narrow-band reduction algorithms
without the conventional tridiagonalization procedure.
The solvers are called
\lq ELPA2' (routine name in the code : \lq solve\_evp\_real\_2stage')
for the ELPA routine
and \lq Eigen\_sx' or \lq EIGX'  (routine name in the code : \lq eigen\_sx')
for the EigenExa routine in this paper.
See the papers ~\cite{EigenExa-PNST, ELPAAlgorithm} for details.

For (b) the reducer,
three routines are found in the base libraries and are called
ScaLAPACK style, ELPA style, and EigenExa style reducers in this paper.
In the ScaLAPACK style,
the Cholesky factorization, Eq.~(\ref{EQ-CHOLE-DECMP})
is carried out and then
the reduced matrix $A'$, defined in Eq.~(\ref{EQ-A-ATAU}),
is generated by a recursive algorithm
(routine name \lq pdsygst')
without explicit calculation of $U^{-1}$ nor $U^{-T}$.
Details of the recursive algorithm are explained,
for example in Ref.~\cite{ReducingGEP}.
In the ELPA style,
the Cholesky factorization
(routine name: \lq cholesky\_real')
is carried out, as in the ScaLAPACK style,
and the reduced matrix $A'$ is generated
by the explicit calculation of
the inverse (triangular) matrix $R\equiv U^{-1}$
(routine names : \lq invert\_trm\_real')
and the explicit successive matrix multiplication of $A' = (R^T A) R$
(routine names: \lq mult\_at\_b\_real')
\cite{ELPAReview}
\footnote{The benchmark was carried out
  in an ELPA style reduction algorithm.
  The ScaLAPACK  routine of \lq pdtrmm'
  is used for the multiplication of the triangular matrix $R$ from right,
  while a sample code in the ELPA package
  uses the ELPA routine (\lq mult\_at\_b\_real').
  We ignore the difference,
  since the elapse time of the above procedure is not dominant.}.
In the EigenExa style,
the Cholesky factorization is not used.
Instead, the SEP for the matrix $B$
\begin{eqnarray}
  B W = W D,
  \label{EQ-SEV-EQ-B-MAT}
\end{eqnarray}
is solved by the SEP solver (Eigen\_sx),
with the diagonal matrix of $D \equiv {\rm diag}(d_1, d_2,...)$
and the unitary matrix of $W \equiv (\bm{w}_1 \, \bm{w}_2 \, ....)$.
A reduced SEP in the form of
Eq.~(\ref{EQ-SEV-EQ}) is obtained by
\begin{eqnarray}
  A' &=&  (D^{-1/2} W^{T}) A (W D^{-1/2}) \\
  Y &=& W  D^{-1/2} Z,
  \label{EQ-SEV-EQ-EIGENEXA}
\end{eqnarray}
because of $Z=D^{1/2} W^{T} Y$ and $W^{-T}=W$.
Equation (\ref{EQ-SEV-EQ-EIGENEXA})
is solved by the SEP solver  (Eigen\_sx).

Though
the SEP solver of Eq.~(\ref{EQ-SEV-EQ}) requires
a larger operation cost than the Cholesky factorization
(See Fig.1 of Ref.~\cite{CHOLESKY-EXP}, for example),
the elapse time can not be estimated only from the operation costs
among the modern supercomputers.

\begin{table}[htb]
  \caption{List of the workflows in the benchmark. The routine names for the SEP solver and the reducer are shown for each workflow. Abbreviations are shown within parentheses.  \label{tab:workflows}
  }
  \begin{tabular}{c|cc}
    Workflow & SEP solver   & Reducer       \\ \hline
    $A$        & ScaLAPACK (SCLA)  & ScaLAPACK (SCLA)  \\
    $B$        & Eigen\_sx (EIGX)  & ScaLAPACK (SCLA)  \\
    $C$        & ScaLAPACK (SCLA)  & ELPA   \\
    $D$        & ELPA2 & ELPA  \\
    $E$        & ELPA1 & ELPA   \\
    $F$        & Eigen\_s (EIGS)  & ELPA   \\
    $G$        & Eigen\_sx (EIGX)  & ELPA   \\
    $H$        & Eigen\_sx (EIGX)  & Eigen\_sx (EIGX)
  \end{tabular}
\end{table}


The benchmark of the hybrid GEP solvers
was carried out
for the eight workflows listed in Table~\ref{tab:workflows}.
In general, a potential issue
is the possible overhead of the data conversion process between libraries.
This issue will be discussed in Sec.~\ref{SEC-DATA-CONV}.

\begin{table}[htb]
  \caption{Selected results of the benchmark.  The elapse time for the full (eigenpair) calculation ($T_{\rm full}$) and that for the eigenvalue-only calculation ($T_{\rm evo}$) with the workflows. The recorded time is the best data among ones with different numbers of the used nodes.  The number of used nodes ($P$) for the best data is shown within parentheses. The best data among the workflows are labelled by \lq [B]'. The saturated data are labelled by \lq [S]'.
    The workflow $D'$ on Altix is that without the SSE optimized routine of the \lq ELPA2' SEP solver.
    See the text for details. }
  \begin{tabular}{cc|c|cc}
    Size $M$/Machine & WF    & $T_{\rm full}$ (sec)   & $T_{\rm evo}$ (sec) \\ \hline \hline
    1,000,080/FX10 & $G$ & 39,919 ($P$ = 4,800) & 35,103 ($P$ = 4,800) \\ \hline \hline
    430,080/K & $A$      & 11,634 ($P$ = 10,000) & 10,755 ($P$ = 10,000) \\
          & $B$          &  8,953 ($P$ = 10,000) &  8,465 ($P$ = 10,000) \\
          & $C$          &  5,415 ($P$ = 10,000) &  4,657 ($P$ = 10,000) \\
          & $D$          &  4,242 ($P$ = 10,000) &  2,227 ($P$ = 10,000)[B] \\
          & $E$          &  2,990 ($P$ = 10,000) &  2,457 ($P$ = 10,000) \\
          & $F$          &  2,809 ($P$ = 10,000) &  2,416 ($P$ = 10,000) \\
          & $G$          &  2,734 ($P$ = 10,000)[B] &  2,355 ($P$ = 10,000) \\
          & $H$          &  3,595 ($P$ = 10,000) &  3,147 ($P$ = 10,000) \\ \hline \hline
    90,000/K & $A$       &    590 ($P$ = 4,096)  &    551 ($P$ = 4,096)  \\
          & $B$          &    493 ($P$ = 1,024)[S]  &    449 ($P$ = 1,024)[S]  \\
          & $C$          &    318 ($P$ = 4,096)  &    298 ($P$ = 4,096)  \\
          & $D$          &    259 ($P$ = 4,096)  &    190 ($P$ = 4,096)[B]  \\
          & $E$          &    229 ($P$ = 4,096)[B]  &    194 ($P$ = 4,096)  \\
          & $F$          &    233 ($P$ = 4,096)  &    210 ($P$ = 4,096)  \\
          & $G$          &    258 ($P$ = 4,096)  &    240 ($P$ = 4,096)  \\
          & $H$          &    253 ($P$=4,096)  &    236 ($P$=4,096)  \\ \hline
    90,000/FX10 & $A$    &  1,248 ($P$ = 1,369)  &  1,183 ($P$ = 1,369)  \\
          & $B$          &    691 ($P$ = 1,024)[S]  &    648 ($P$ = 1,024)[S]  \\
          & $C$          &    835 ($P$ = 1,369)  &    779 ($P$ = 1,369)  \\
          & $D$          &    339 ($P$ = 1,369)  &    166 ($P$ = 1,024)[B][S]  \\
          & $E$          &    262 ($P$ = 1,369)  &    233 ($P$ = 1,024)[S]  \\
          & $F$          &    250 ($P$ = 1,369)[B]  &    222 ($P$ = 1,369)  \\
          & $G$          &    314 ($P$ = 1,024)[S]  &    283 ($P$ = 1,024)[S]  \\
          & $H$          &    484 ($P$=1,369)  &    456 ($P$=1,369)  \\ \hline
    90,000/Altix & $A$   &  1,985 ($P$ = 256)    &  1,675 ($P$ = 256)    \\
          & $B$          &  1,883 ($P$ = 256)    &  1,586 ($P$ = 256)    \\
          & $C$          &  1,538 ($P$ = 256)    &  1,240 ($P$ = 256)    \\
          & $D$          &  1,621  ($P$ = 256)   &     594 ($P$ = 256)  \\
          & $D'$         &  2,621  ($P$ = 256)   &     585 ($P$ = 256)[B]  \\
          & $E$          &  1,558 ($P$ = 256)    &  1,287 ($P$ = 256)    \\
          & $F$          &  1,670 ($P$ = 256)    &  1,392 ($P$ = 256)    \\
          & $G$          &  1,453 ($P$ = 256)[B]    &  1,170 ($P$ = 256)    \\
          & $H$          &    2,612 ($P$=256)  &    2,261 ($P$=256)  \\ \hline \hline
    22,500/K & $A$       &   65.2 ($P$ = 1,024)  &   59.6 ($P$ = 256)    \\
          & $B$          &   45.8 ($P$ = 1,024)[S]  &   43.2 ($P$ = 1,024)[S]  \\
          & $C$          &   41.7 ($P$ = 2,025)  &   37.8 ($P$ = 2,025)  \\
          & $D$          &   28.4 ($P$ = 2,025)  &   22.6 ($P$ = 1,024)  \\
          & $E$          &   28.3 ($P$ = 2,025)[B]  &   22.6 ($P$ = 1,024)[B]  \\
          & $F$          &   28.8 ($P$ = 1,024)[S] &   26.9 ($P$ = 1,024)[S]  \\
          & $G$          &   29.7 ($P$ = 1,024)[S]  &   27.8 ($P$ = 1,024)[S]  \\
          & $H$          &    39.3($P$=1024)[S]  &    37.5($P$=1024)[S]  \\ \hline
    22,500/FX10 & $A$    &   126.2 ($P$ = 256)   &   118.1 ($P$ = 256)   \\
          & $B$          &    71.3 ($P$ = 256)[S]   &    67.1 ($P$ = 256)[S]   \\
          & $C$          &   103.5 ($P$ = 256)[S]   &    96.3 ($P$ = 256)[S]   \\
          & $D$          &    30.5 ($P$ = 529)[B]   &    24.4 ($P$ = 529)[B]   \\
          & $E$          &    34.3 ($P$ = 256)[S]   &    31.2 ($P$ = 256)[S]   \\
          & $F$          &    32.1 ($P$ = 529)   &    29.4 ($P$ = 529)   \\
          & $G$          &    45.3 ($P$ = 529)   &    42.5 ($P$ = 529)   \\
          & $H$          &    74.9($P$=529)  &    72.2 ($P$=529)  \\ \hline
    22,500/Altix & $A$   &    51.4 ($P$ = 256)   &    42.1 ($P$ = 256)   \\
          & $B$          &    70.0 ($P$ = 256)   &    50.7 ($P$ = 256)   \\
          & $C$          &    45.6 ($P$ = 256)   &    35.5 ($P$ = 256)   \\
          & $D$          &    41.8  ($P$ = 256)  &    22.3  ($P$ = 256)[B]   \\
          & $D'$         &    59.6  ($P$ = 256)  &    21.8  ($P$ = 256)[B]   \\
          & $E$          &    32.3 ($P$ = 256)[B]   &  26.7 ($P$ = 256)   \\
          & $F$          &    48.5 ($P$ = 256)   &    37.3 ($P$ = 256)   \\
          & $G$          &    57.2 ($P$ = 256)   &    39.6 ($P$ = 256) \\
          & $H$          &    71.2 ($P$=256)  &    64.1 ($P$=256)
  \end{tabular}
  \label{tab:besttime}
\end{table}

\section{Benchmark result}
\label{SEC-BENCHMARK}

Strong scaling  benchmarks
are investigated for the hybrid solvers.
The elapse times were measured for
(i) the full eigenpair calculation ($T_{\rm full}$)
and (ii) the \lq eigenvalue-only' calculation ($T_{\rm evo}$).
In the latter case,
the elapse time is ignored for the calculation of
the eigenvectors.
The two types of calculations are
important among electronic structure calculations.
\cite{ELPAReview}
The present benchmark ignores
small elapse times of
the initial procedure for distributed data and
the comments on them will appear in  Sec.~\ref{SEC-INI-DATA}.

The benchmark was carried out on three supercomputers; the K computer at Riken,  Fujitsu FX10 and SGI Altix ICE 8400EX.  The K computer has a single SPARC 64 VIIIfx processor (2.0GHz, 8-core) on node. The FX10 is Oakleaf-FX of the University of Tokyo. Fujitsu FX10 is the successor of the K computer and has a single SPARC64 IXfx processor (1.848 GHz, 16-core) on each node.
\footnote{
  Additional options of the K computer and FX10 are explained;
  We did not specify a MPI process shape on the Tofu interconnect.
  We used the rank directory feature to alleviate I/O contention.
}
We also used SGI Altix ICE 8400EX of Institute for Solid State Physics of the University of Tokyo.
It is a cluster of Intel Xeon X5570 (2.93GHz, 8-core).
The byte-per-flop value (B/F) is B/F=0.5, 0.36 or 0.68, for the K computer, FX10 or SGI Altix, respectively.
The numbers of used processor nodes $P$ are
set to be square numbers ($P=q^2$)
except in Sec.~\ref{SEC-BENCH-MILLION},
since the ELPA paper~\cite{ELPAReview} reported
that the choice of a (near-)square number for $P$
can give better performance.

When the non-traditional SEP solver algorithm of ELPA is used on Altix,
one can choose
an optimized low-level routine using SSE instructions
(\lq REAL\_ELPA\_KERNEL\_SSE')
and a generic routine
(\lq REAL\_ELPA\_KERNEL\_GENERIC'). \cite{ELPAReview}
The optimized code can run only on the Intel-based architectures
compatible to SSE instructions
and was prepared so as to accelerate the backtransformation subroutine.
Among the results on Altix,
the \lq ELPA2' solver and the workflow $D$ on Altix
are those with the optimized routine,
while the \lq ELPA2$'$' solver and the workflow $D'$
are those with the generic routine.

\begin{figure}[htb]
  \centering
   \includegraphics[width=8.5cm]{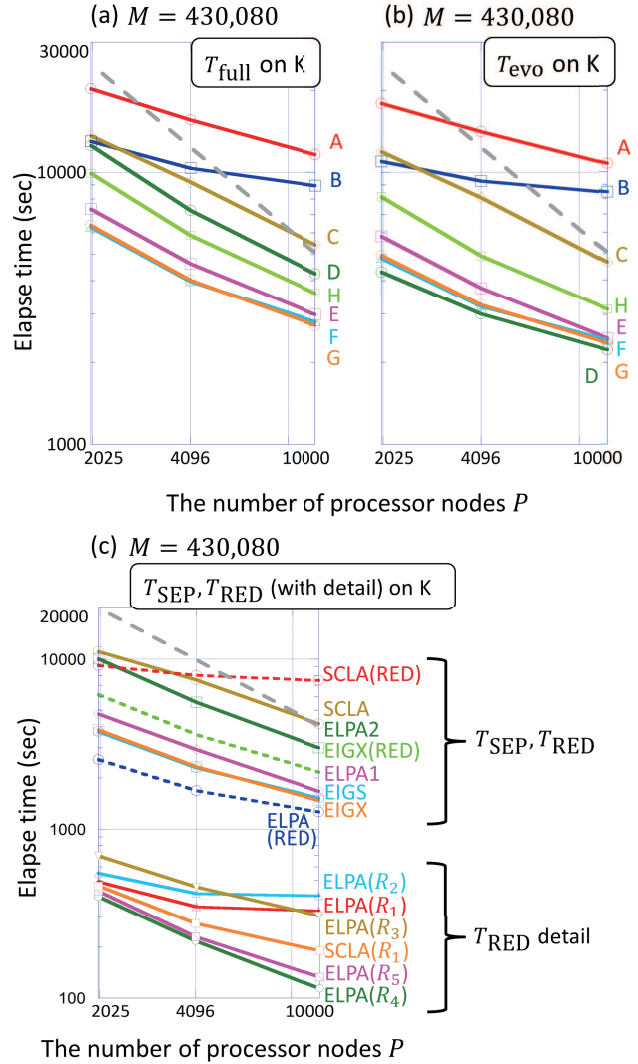}
  \caption{
    Results with $M$=430,800 on the K computer.
    The elapse times are plotted with the workflows
    for the (a) full ($T_{\rm full}$) and (b) eigenvalue-only ($T_{\rm evo}$) calculations.
    (c) The decomposed times for the SEP solver ($T_{\rm SEP}$)
    and for the reducer ($T_{\rm RED}$) are plotted.
    The routines for the reducers is labeled by \lq (RED)'.
    Detailed decomposed times for subprocedures of the ELPA style reducer and the Cholesky decomposition in the ScaLAPACK style reducer are also plotted in (c).
    The ideal speedup in parallelism is drawn as a dashed gray line.
  }
  \label{fig:combined_430080_K}
\end{figure}


\subsection{Result with the matrix size of $M=430,080$}
\label{subsec:430080}

The benchmark with the matrix size of $M=430,080$ was carried out
for up to $P$ = 10,000 nodes on the K computer.
The elapse times for $P$=10,000 nodes
is shown in Table~\ref{tab:besttime}.
The elapse time for all the cases are shown
in Fig.~\ref{fig:combined_430080_K}
for the (a) full ($T_{\rm full}$) or (b) eigenvalue-only ($T_{\rm evo}$) calculations.
The decomposed times are also shown in
Fig.~\ref{fig:combined_430080_K} (c) for the SEP solver ($T_{\rm SEP}$)
and the reducer ($T_{\rm RED}$)
($T_{\rm full} = T_{\rm SEP} + T_{\rm RED}$).

\begin{table}[htb]
  \caption{Decomposition of the elapse time (sec) of the SEP solvers with $M$=430,080 and $P=10,000$. See the text for the subroutine names of \lq TRD/BAND' , \lq D\&C' and \lq BACK'. }
  \begin{tabular}{c|ccc|c}
    SEP solver & TRD/BAND & D\&C & BACK & Total ($T_{\rm SEP}$) \\ \hline
    SCLA  & 3,055 & 465 &  633 & 4,152 \\
    ELPA2 &  966 & 141 & 1,892 & 2,999 \\
    ELPA1 & 1,129 & 138 &  400 & 1,667 \\
    EIGS  & 1,058 & 196 &  265 & 1,521 \\
    EIGX  &  828 & 390 &  255 & 1,473
  \end{tabular}
  \label{tab:SEP_decomposition}
\end{table}

Table~\ref{tab:SEP_decomposition} shows
the decomposed time of the SEP solvers for $P$=10,000.
A SEP solver routine is decomposed into three subroutines of
(i) the tridiagonalization or narrow-band reduction (\lq TRD/BAND'),
(ii) the divide and conquer algorithms
for the tridiagonal or narrow-band matrices (\lq D\&C')
so as to compute the eigenvalues, and
(iii) the backtransformation of eigenvectors (\lq BACK')
so as to compute the eigenvectors of the GEP.

One can observe several features on the results;
(I) In the full calculation benchmark (Fig.~\ref{fig:combined_430080_K}(a)),
the best data, the smallest elapse time, appears
in the workflow $G$ for $P$=10,000.
The workflow $G$ is the hybrid solver that uses
the \lq Eigen\_sx' SEP solver in EigenExa and
the ELPA style reducer,
since these routines are the best
among the SEP solvers and the reducers, respectively,
as shown in
Fig.~\ref{fig:combined_430080_K}(c) and Table~\ref{tab:SEP_decomposition}.
In Table~\ref{tab:besttime},
the speed ($T_{\rm full}^{-1}$) of the workflow $G$
is approximately four times faster than
that of the conventional workflow $A$
(11,634 sec) / (2,734 sec)  $\approx$ 4.3).
(II)
Fig.~\ref{fig:combined_430080_K} (c) shows that
the ELPA style reducer gives
significantly smaller elapse times than
those of ScaLAPACK and those of EigenExa.
The elapse time  for $P$=10,000
is $T_{\rm RED}$ = 1,261 sec with the ELPA style reducer
and is $T_{\rm RED}$ = 2,157 sec with the EigenExa reducer.
The elapse time with
the EigenExa reducer is governed by
that of the SEP solver for Eq.~(\ref{EQ-SEV-EQ-B-MAT})
($T_{\rm SEP}$ = 1,473 sec in
Table~\ref{tab:SEP_decomposition}).
(III)
In the eigenvalue-only calculation
(Fig.~\ref{fig:combined_430080_K}(b)),
the best data, the smallest elapse time, appears
in the workflow $D$ for $P$=10,000.
The workflow $D$ is the solver that uses
the \lq ELPA2' SEP solver and
the ELPA style reducer and
the eigenvector calculation consumes
a large elapse time of $T_{\rm vec}$;
$T_{\rm vec} \equiv T_{\rm full} - T_{\rm evo}
=$ (4,242 sec) - (2,227 sec) = (2,015 sec)
in Table~\ref{tab:besttime}.
The time $T_{\rm vec}$ is contributed mainly
by the backward transformation subroutine
($T_{\rm BACK}$ =1,892 sec)
in Table~\ref{tab:SEP_decomposition},
because the backward transformation subroutine in ELPA2
uses a characteristic two-step algorithm
(See Sec. 4.3 of Ref.~\cite{ELPAReview}).

\subsection{Benchmark with the matrix sizes of
  $M$=90,000, 22,500 \label{SEC-RESULT-MIDDLESIZE}}

\begin{figure}[h]
  \centering
  \includegraphics[width=8.5cm]{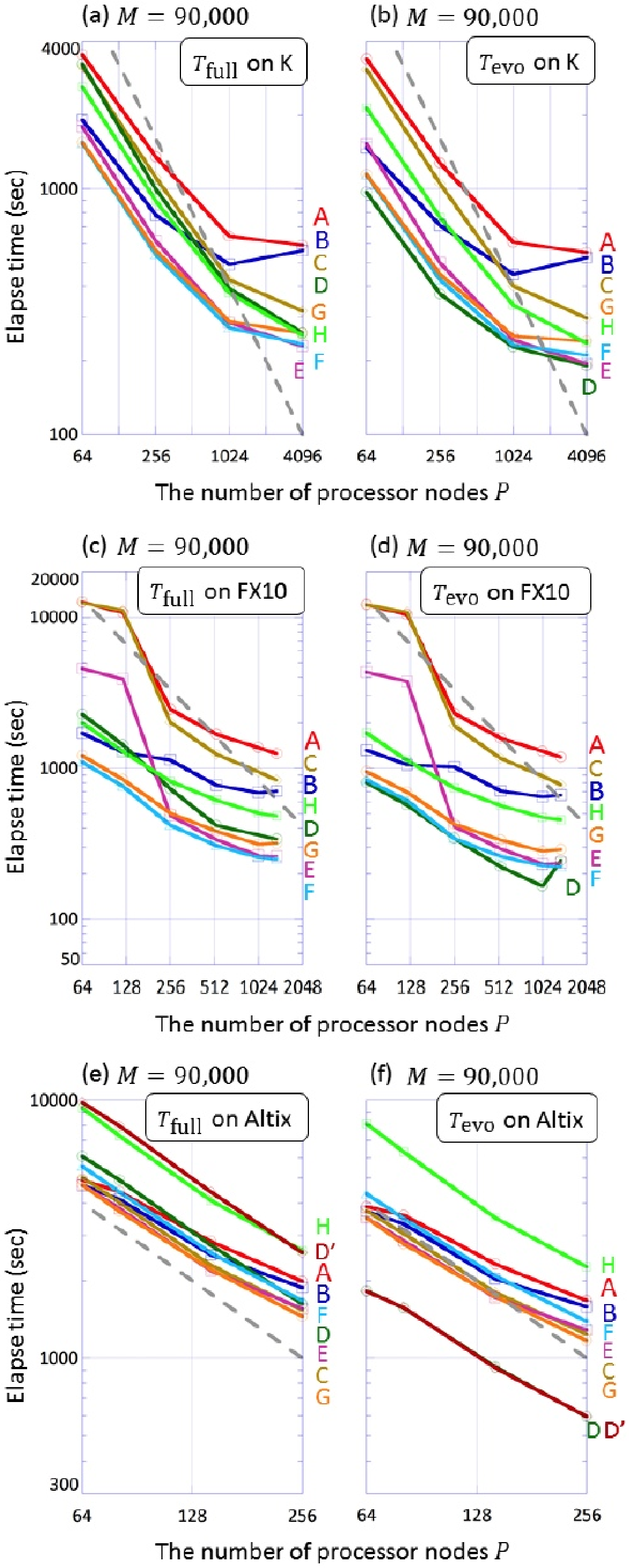}
  \caption{Benchmark with the matrix size of M=90,000, (I) on the K computer for the (a) full (eigenpair) and (b) eigenvalue-only calculation, (II) on FX10 for  the (c) full  and (d) eigenvalue-only calculation, (III) on Altix for  the (e) full  and (f) eigenvalue-only calculation. The ideal speedup in parallelism is drawn  as a dashed gray line.
    \label{fig:combined_90000_all_machines}}
\end{figure}

\begin{figure}[h]
  \centering
  \includegraphics[width=8.2cm]{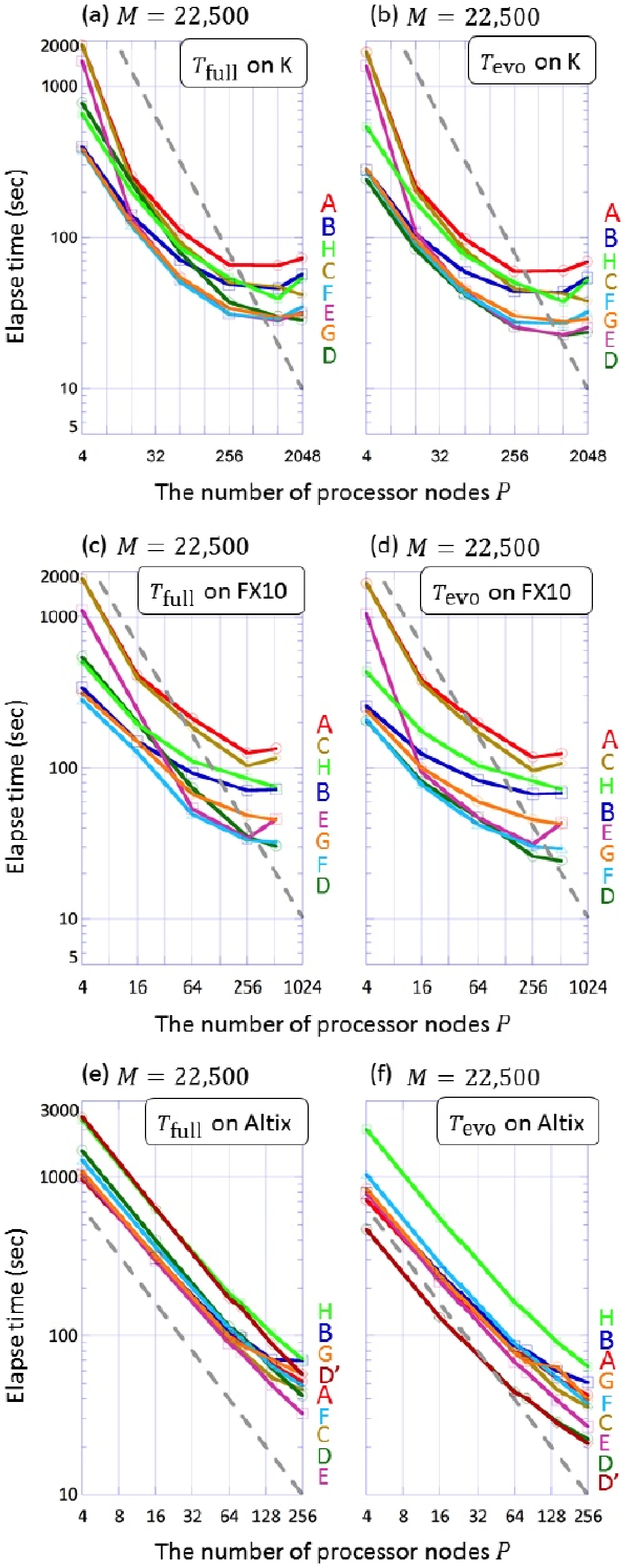}
  \caption{Benchmark with the matrix size of M=22,500, (I) on the K computer for the (a) full (eigenpair) and (b) eigenvalue-only calculation, (II) on FX10 for  the (c) full  and (d) eigenvalue-only calculation, (III) on Altix for  the (e) full  and (f) eigenvalue-only calculation. The ideal speedup in parallelism is drawn  as a dashed gray line.
    \label{fig:combined_22500_all_machines}}
\end{figure}

\begin{figure}[h]
  \centering
  \includegraphics[width=8.5cm]{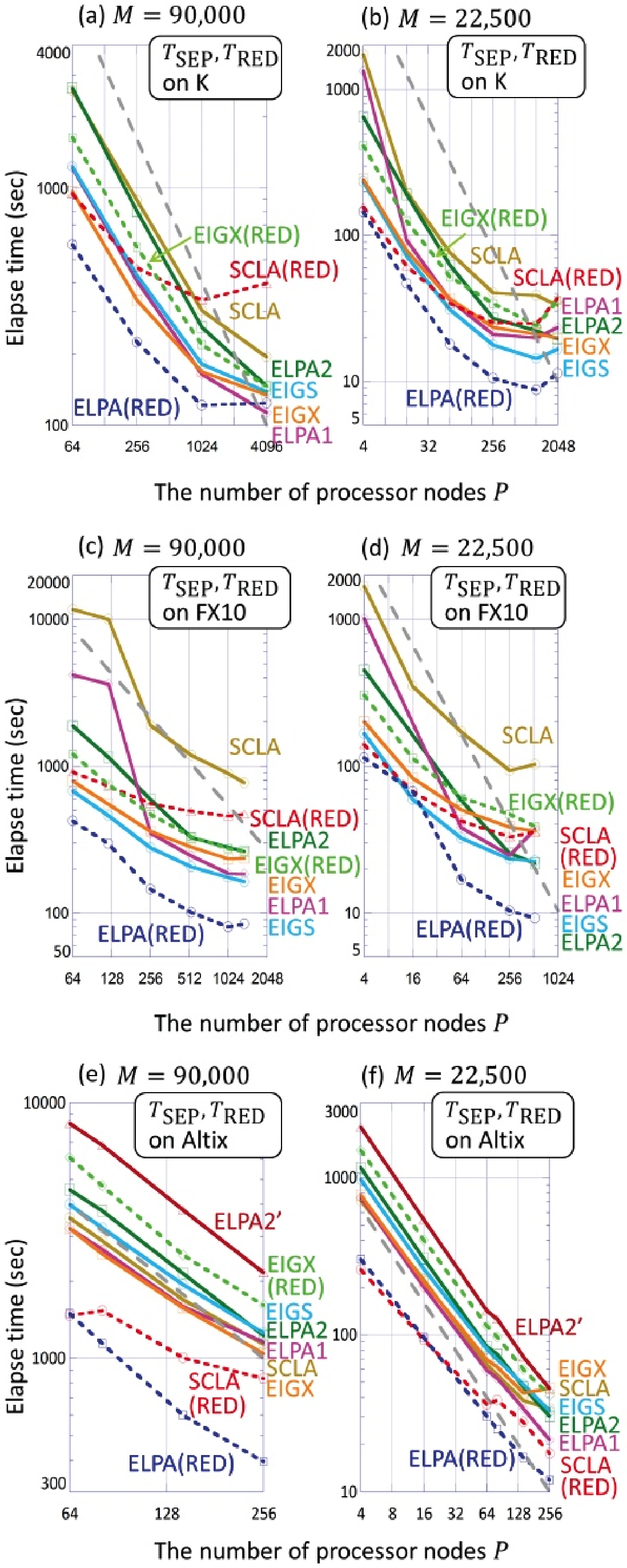}
  \caption{Decomposition analysis of the elapse time into those of the SEP solver and the reducer (I) on the K computer with (a) $M$=90,000 and (b) $M$=22,500, (II) on FX10 with (c) $M$=90,000 and (d) $M$=22,500, (III) on Altix with (c) $M$=90,000 and (d) $M$=22,500.
    The routines for the reducers is labeled by \lq (RED)'.
    The \lq ELPA2$'$' SEP solver is that
    without the SSE optimized routine.
    The ideal speedup in parallelism is drawn  as a dashed gray line.\label{fig:combined_seprep_all_machines}}
\end{figure}

The benchmark with the smaller matrix sizes of $M=90,000$ and 22,500
are also investigated.
The maximum number of used processor nodes
is $P_{\rm max}$ =  4,096, 1,039 and 256,
on the K computer, FX10, and Altix, respectively.
\footnote{
  We observed on Altix that
  the \lq ELPA2'
  and \lq ELPA2$'$' SEP
  solver required
  non-blocking communication requests
  beyond the default limit number
  of $N_{\rm MPI\_MAX}=16,384$
  and the job stopped with an MPI  error message.
  Then we increased the limit number
  to $N_{\rm MPI\_MAX}=1,048,576$,
  the possible maximum of the machine
  by the environment variable
  \lq MPI\_REQUEST\_MAX'
  and the calculations were completed.
}
Figures~\ref{fig:combined_90000_all_machines}
and ~\ref{fig:combined_22500_all_machines}
show the data with $M$=90,000 and with $M$=22,500, respectively.
The decomposed times are shown in
Fig.~\ref{fig:combined_seprep_all_machines}.
Table \ref{tab:besttime} shows
the best data for each workflow among the different numbers of used nodes.
The results will help general simulation researchers
to choose the solver and the number of used nodes,
since the elapse times in Table~\ref{tab:besttime}
are less than a half hour and
such calculations are popular \lq regular class' jobs
among systematic investigations.
\footnote{
  One should remember that supercomputers
  are usually shared by many researchers
  who run many calculations in similar problem sizes
  successively and/or simultaneously.
}

Here, the results  are discussed;
(I) Table~\ref{tab:besttime}  shows that
the smallest elapse time in the full calculation
appears among the workflows with the ELPA style reducer
(the workflows $D, E, F$, and $G$) and
that in the eigenvalue-only calculation
appears with the workflow $D$.
The above features are consistent
to the results in the previous subsection.
(II) Unlike the result in the previous subsection,
the speed up is sometimes saturated.
An example is observed
in Fig.~\ref{fig:combined_22500_all_machines} (a),
in the full calculation with $M$=22,500 on the K computer,
because the elapse time in the workflow $F$
gives a minimum as the function of $P$ at $P$=1,024.
The decomposition analysis of
Fig.~\ref{fig:combined_seprep_all_machines}(b) indicates that
the saturation occur both
for the SEP solver and the reducer,
which implies that
the improvement both on the SEP solver and the reducer is desirable.
The saturated cases are marked in Table~\ref{tab:besttime}
with the label of  \lq [S]'.
\footnote{
  No saturation is found on Altix,
  unlike on the K computer and FX10,
  partially because
  the maximum number of used nodes
  ($P_{\rm max}=256$) is smaller.
}
(III) Finally,
the  SSE-optimized routine in the workflow $D$
is compared with the generic routine in the workflow $D'$
in the case of $M=90,000$ on Altix with $P=$256.
The SSE-optimized routine is prepared only
in the backward transformation process.
Since the process with the SSE-optimized routine or the generic one
gives the elapse time of $T_{BACK}$ = 929 sec or
$T_{BACK}$ = 1,872 sec, respectively,
the process is accelerated
with the SSE-optimized routine
by  $1,872$ sec / $929$ sec $\approx$ 2.02.
As shown in Table~\ref{tab:besttime},
the full calculation is accelerated
with the SSE-optimized routine
by  $2,621$ sec / $1,621$ sec $\approx$ 1.62.



\subsection{Benchmark for a million dimensional matrix \label{SEC-BENCH-MILLION}}

Finally, the benchmark for a million dimensional matrix is discussed.
A press release at 2013 \cite{EigenExaPressRelease} reported, as a world record,
a benchmark of a million dimensional SEP carried out by EigenExa,
in approximately one hour,
on the full (82,944) nodes of the K computer.
An eigenvalue problem with a million dimensional matrix ($M$=10$^6$)
seems to be the practical limitation of the present supercomputer,
owing to the $\mathcal{O}(M^3)$ operation cost.

We calculated a million dimensional GEP at Dec. 2014
on the full (4,800) nodes of Oakleaf-FX.
\footnote{
  We used FX10 not the K computer,
  because FX10 is in a newer architecture with a lower B/F value
  and the result on FX10 is speculated
  to be closer to that on the next-generation (exa-scale) machine.
}
Since our computational resource was limited,
only one calculation was carried out
with the workflow $G$,
because it gives the best data
among those with $M=430,080$ in Table~\ref{tab:besttime}.
The calculation finished in approximately a half day,
as shown in Table~\ref{tab:besttime}
($T_{\rm full}$ = 39,919 sec and $T_{\rm evo}$ = 35,103 sec).
The elapse time of the reducer
($T_{\rm RED}=T_{\rm full} - T_{\rm SEP}$ = 15,179 sec)
is smaller than but comparable to that of the SEP solver
($T_{\rm SEP}=24,740$).
The benchmark proved that
the present code
qualifies as a software
applicable to massively parallel computation
with up to a million dimensional matrix.

\section{Discussions}
\label{SEC-DISCUSSIONS}

\subsection{Preparation of initial distributed data \label{SEC-INI-DATA}}

In the benchmark,
the initial procedures including file reading
are carried out for the preparation of distributed data.
Its elapse time is always small and is ignored in the previous section.
\footnote{
In the case of
the workflow G on the K computer
with $M$=430,080 and $P$=10,000,
for example,  the elapse time of the initial procedures
is $T_{\rm ini}=$123sec and is much smaller
than that of the total computation
($T_{\rm tot}=2,734$sec. See Table. \ref{tab:besttime}).
It is noteworthy that the present matrices are sparse,
as explained in Sec.~\ref{SEC-BACKGROUND}.
}
These procedures, however,
may consume significant elapse times,
when the present solver is used as a middleware with real applications.
The discussions on such cases are beyond the present scope,
since they depend on the program structure of the real applications.
Here, several comments are added
for real application developers;
In general,
the matrix data cost is, at most,
$\mathcal{O}(M^2)$
and the operation cost is $\mathcal{O}(M^3)$ in the dense-matrix solvers and
one should consider a balance between them.
In the case of $M=430,080$, for example,
the required memory size for all the matrix elements
is 8 B $\times \, M^2 \approx$ 1.5 TB,
which can not be stored on a node of the K computer.
Therefore, the data should be always distributed.
In our real application (ELSES),
the initial distributed data is prepared,
when only the required elements are generated and stored on each node.

\subsection{Data conversion overhead \label{SEC-DATA-CONV}}

As explained in Sec.~\ref{SEC-MATH-FOUND},
several workflows require data conversion processes between
distributed data formats,
since
ScaLAPACK and ELPA use block cyclic distribution
with a given block size $n_{\rm block}(>1)$ and
EigenExa uses
cyclic distribution ($n_{\rm block} \equiv 1$).
In the present benchmark,
the block size $n_{\rm block}$ in ScaLAPACK and ELPA
was set to be $n_{\rm block}=128$, a typical value.
Consequently,
the workflows B, F, G require data conversion processes.
In the present paper,
the elapse time of the conversion procedures is included
in the reducer part ($T_{\rm red}$).

Table~\ref{tab:conversion} shows the elapse time
for the data conversion.
The elapse times are shown in the cases with the maximum numbers of used nodes ($P=P_{\rm max}$)
among the present benchmark.
Two data conversion procedures are required.
One is the conversion from the block cyclic distribution into the cyclic distribution,
shown as \lq (b $\rightarrow$ 1)' in Table~\ref{tab:conversion} and
the other is the inverse process shown as \lq (1 $\rightarrow$ b)'.
The two procedures are carried out, commonly, by the \lq pdgemr2d' routine in ScaLAPACK.

Table~\ref{tab:conversion} indicates that
the overhead of the data conversion procedures is always small
and is not the origin of the saturation.
In general,
the conversion requires an $\mathcal{O}(M^2)$ operation cost,
while the calculation in a dense-matrix solver
requires an $\mathcal{O}(M^3)$ operation cost.
The fact implies the general efficiency of hybrid solvers,
at least, among dense-matrix solvers.

\begin{table}[htb]
  \caption{The elapse times for data conversion; \lq (b $\rightarrow$ 1)', \lq (1 $\rightarrow$ b)' and
  \lq $T_{\rm RED}$' are the times in seconds for, the conversion process from block cyclic into cyclic distributions, the inverse process and
  the whole reducer procedure, respectively. The saturated data of $T_{\rm RED}$ are labelled by \lq [S]'. The \lq ratio' is
  ((b $\rightarrow$ 1) + (1 $\rightarrow$ b)) / $T_{\rm RED}$.}
  \label{tab:conversion}
  \begin{tabular}{cc|ccc|c}
      Size M & Machine(P)    &   (b $\rightarrow$ 1) &  (1 $\rightarrow$ b) & $T_{\rm RED}$ & ratio[\%] \\
      \hline
      1,008,000 & FX10(4,800) & 51.4    & 51.7   & 8,208   & 1.26 \\
      430,080 & K(10,000)   & 13.4    &  6.48  & 1,261   & 1.58 \\
      90,000 & K(4,096)     &  6.89   &  0.797 & 124[S]  & 6.21 \\
             & FX10(1,369)  &  1.89   &  0.973 & 84.0[S] & 3.41 \\
             & Altix(256)   &  2.01   &  2.02    & 394     & 1.02 \\
      22,500 & K(2,025)     &  0.571  &  0.610 & 11.3[S] & 10.4 \\
             & FX10(529)    &  0.328  &  0.176 & 9.20    & 5.48 \\
             & Altix(256)   &  0.120  &  0.279   & 11.9    & 3.35 \\
  \end{tabular}
\end{table}

\subsection{Decomposition analysis of the reducer}

\begin{figure}[htb]
  \centering
  \includegraphics[width=8.5cm]{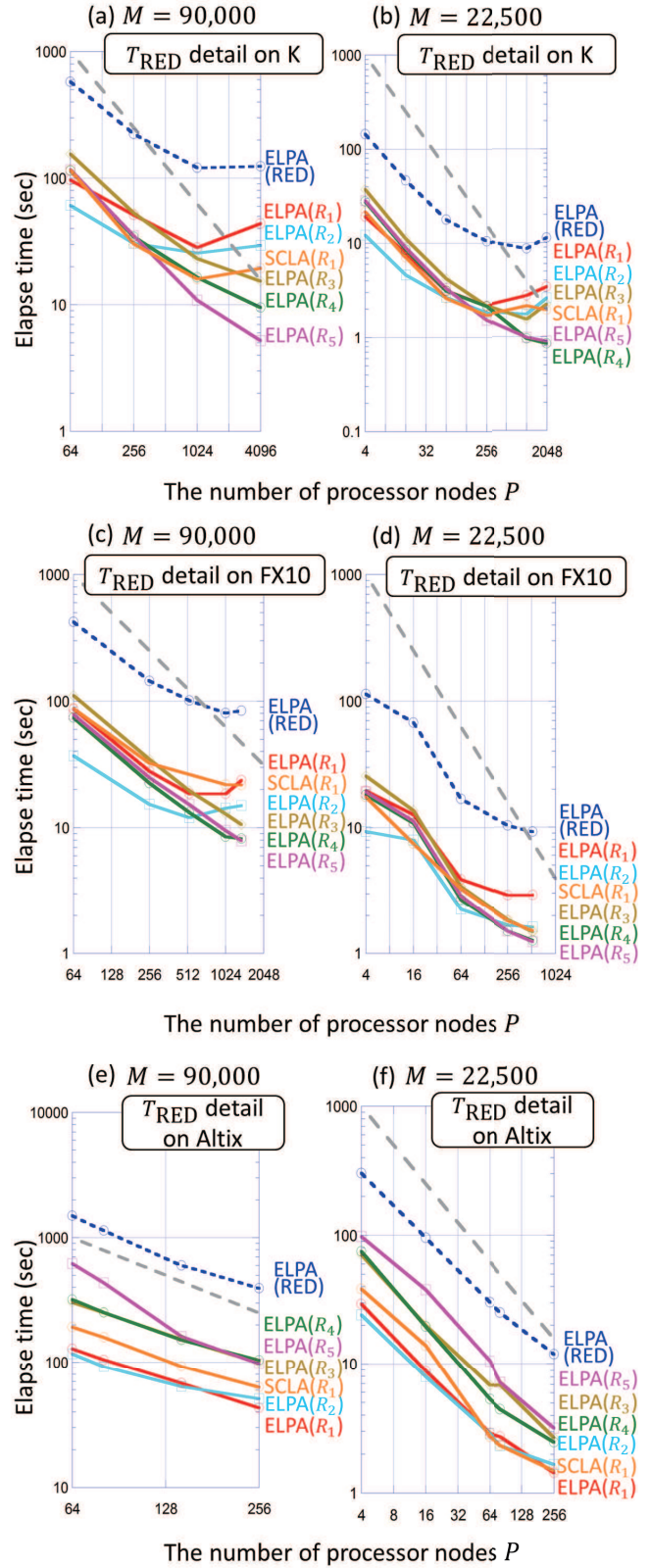}
  \caption{Decomposition analysis of the elapse time of subprocedures of the ELPA style reducer and the Cholesky factorization in the ScaLAPACK style reducer (I) on the K computer with (a) $M$=90,000 and (b) $M$=22,500, (II) on FX10 with (c) $M$=90,000 and (d) $M$=22,500, (III) on Altix with (c) $M$=90,000 and (d) $M$=22,500.
    The ideal speedup in parallelism is drawn as a dashed gray line.}
  \label{fig:ReducerDetail}
\end{figure}

The decomposition analysis of the ELPA-style reducer is focused on,
since the ELPA-style reducer is fastest among the three libraries.
Figure~\ref{fig:combined_430080_K} (c) shows the case on the K computer with $M$=430,080.
The elapse times of the subprocedures of the ELPA-style reducer are plotted;
\lq ELPA($R_1$)' is the Cholesky factorization of Eq.~(\ref{EQ-CHOLE-DECMP}), \lq ELPA($R_2$)' is the explicit calculation of the inversion $R = U^{-1}$ of the Cholesky factor $U$, \lq ELPA($R_3$)' and \lq ELPA($R_4$)' are the successive matrix multiplication of Eq.~(\ref{EQ-A-ATAU}) and \lq ELPA($R_5$)' is the backward transformation of eigenvectors by matrix multiplication of Eq.~(\ref{EQ-BACKWARD-TRANS}). The elapse times of the Cholesky factorization in the ScaLAPACK style reducer is also plotted as \lq SCLA($R_1$)' as a reference data.
The same decomposition analysis is carried out also for other cases,
as shown in Fig.~\ref{fig:ReducerDetail}.
One can observe that the Cholesky factorization of the ELPA-style reducer does not scale and
sometimes is slower than that of the ScaLAPACK reducer.
In particular,
the saturation of the ELPA-style reducer is caused
by that of the Cholesky factorization
in Fig.~\ref{fig:ReducerDetail} (a)(b)(c).

The above observation implies that
the reducer can be a serious bottleneck in
the next-generation (exa-scale) supercomputers,
though not in the present benchmark.
One possible strategy is the improvement
on the Cholesky factorization for better scalability
and another is the development of
a reducer without the Cholesky factorization,
as in the EigenExa-style reducer.

\section{Summary and future outlook \label{SEC-SUMMARY}}
In summary,
hybrid GEP solvers were constructed between
the three parallel dense-matrix solver libraries of
ScaLAPACK, ELPA and EigenExa.
The benchmark was carried out
with up to a million dimensional matrix
on the K computer and other supercomputers.
The hybrid solvers with ELPA and EigenExa give
better benchmark results than the conventional ScaLAPACK library.
The code was developed as
a middleware and a mini-application
and will appear online.
Several issues are discussed.
In particular,
the decomposition analysis of the elapse time
reveals a potential bottleneck part
on next-generation (exa-scale) supercomputers,
which indicates the guidance for future development
of the algorithms and the codes.

As a future outlook,
the present code for the hybrid solvers
is planned to be extended
by introducing
the solvers with different mathematical foundations.
A candidate is
the parallel block Jacobi solver
\cite{BLOCK-YACOBI-2012, BLOCK-YACOBI-2014}.
Since the solver is applicable only to standard eigenvalue problems,
the hybrid solver
enables us to use the solver in generalized eigenvalue problems.

\begin{acknowledgment}
  The authors thank to Toshiyuki Imamura and Takeshi Fukaya in RIKEN Advanced
  Institute of Computational Science (AICS) for fruitful discussions on EigenExa.
  The authors also thank to Yusaku Yamamoto
  in The University of Electro-Communications on the parallel block Jacobi solver.
  This
  research is partially supported by Grant-in-Aid for Scientific Research
  (KAKENHI Nos. 25104718 and 26400318) from the Ministry of Education,
  Culture, Sports, Science and Technology (MEXT) of Japan. The K computer
  of RIKEN was used in the research projects of hp140069,
  hp140218, hp150144. The supercomputer Oakleaf-FX of the University of Tokyo was used in
  the research project of 14-NA04 in \lq Joint Usage/Research Center for
  Interdisciplinary Large-scale Information Infrastructures'  in Japan,
  in the \lq Large-scale HPC Challenge' Project, Information Technology
  Center, The University of Tokyo and
  Initiative on Promotion of Supercomputing for Young or Women Researchers, Supercomputing Division, Information Technology Center, The University of Tokyo.
  We also used
  the supercomputer SGI altix ICE 8400EX at the
  Institute for Solid State Physics of the University of Tokyo and
  the supercomputers
  at the Research Center for Computational Science, Okazaki.
\end{acknowledgment}

\end{document}